\documentclass{amsart}
\usepackage{mathrsfs,amssymb,amsmath,amsfonts,amsxtra}
\usepackage{graphicx,color}

\begin{document}

\title{Overlaps in pilot wave field theories}

\begin{abstract}
Recently doubts have been raised about the ability of pilot wave theories with field ontology to recover the predictions of quantum field theory. In particular, Struyve has questioned that the overlap between wave functionals of macroscopically different states with fixed particle number is really non-significant.

With numerical computations and some further plausibility arguments we show that the overlap between n-particle states in field theory decreases almost exponentially with the number of particles and becomes non-significant already for small particle numbers.
\end{abstract}

\author{I. Schmelzer}
\thanks{Berlin, Germany}
\email{\href{mailto:ilja.schmelzer@gmail.com}{ilja.schmelzer@gmail.com}}%
\urladdr{\href{http://ilja-schmelzer.de}{ilja-schmelzer.de}}

\maketitle

\sloppypar \sloppy
\newcommand{\pd}{\partial} %%
\renewcommand{\d}{\mathrm{d}} %%
\renewcommand{\c}{\mbox{$\mathrm{const}$}}
\renewcommand{\Im}{\mbox{$\mathfrak{Im}$}}
\newcommand{\vol}[1]{\mbox{$A_{#1}$}}

\newcommand{\B}{\mbox{$\mathbb{Z}_2$}} 
\newcommand{\Z}{\mbox{$\mathbb{Z}$}}
\newcommand{\R}{\mbox{$\mathbb{R}$}}
\newcommand{\C}{\mbox{$\mathbb{C}$}}
\renewcommand{\H}{\mbox{$\mathcal{H}$}} %%
\renewcommand{\L}{\mbox{$\mathcal{L}^2$}} %%
\newcommand{\T}{\mbox{$\mathcal{T}$}} %%185.91
\newcommand{\V}{\mbox{$\mathcal{V}$}} %%

\newtheorem{theorem}{Theorem}
\newtheorem{criterion}{Criterion}

\providecommand{\abs}[1]{\lvert#1\rvert}
\providecommand{\loo}[1]{\lVert#1\rVert_\infty}
\providecommand{\ltwo}[1]{\lVert#1\rVert_2}

\section{Introduction}

There seems to be no shortage of papers ``proving'' empirical contradictions between de Broglie-Bohm pilot wave theory and quantum theory. Most of them simply suggest that the author does not know elementary facts about pilot wave measurement theory. In particular, Zirpel \cite{Zirpel} (misguided by Streater's ``Lost causes'' \cite{Streater}), followed Streater's own lost cause of proving that pilot wave theory contradicts quantum theory: His argumentation depends on the assumption that the correlations between positions $q(t)$ at different times are observable. They are not, and therefore these correlations cannot be in contradiction with empirical predictions of quantum theory. Similarly, Lapiedra \cite{Lapiedra} does not take into account how multi-time measurements have to be described in pilot wave theory: One has to include the storage devices into the consideration. Then, multi-time measurements will be reduced to a one-time measurement at the end of the experiment, and the one-time equivalence proof applies.

But there have been also some more serious questions about the ability of pilot wave theory to recover quantum predictions in the case of field theory. An interesting issue has been raised, in particular, by Wallace:
\begin{quote}
``\ldots it is crucial for these strategies that they are compatible with decoherence: that is, that the preferred observable is also decoherence-preferred. \ldots a hidden-variable theory whose hidden-variables are not decoherence-preferred will fail \ldots to recover effective quasiclassical dynamics. And in QFT (at least where fermions are concerned) the pointer-basis states are states of definite particle number, which in general are not diagonal in the field observables.''
\cite{Wallace}
\end{quote}
While we do not see sufficient evidence for his thesis, the connection between pilot wave beables and decoherence is not sufficiently clear and deserves some further research.

The aim of this paper is to solve another problem which really causes doubt about the ability of pilot wave field theories to recover quantum preditions. It is the question if the overlap between states with large but fixed particle numbers is non-significant. This problem has been raised by Struyve \cite{Struyve}:
\begin{quote}
Note that the property that macroscopically distinct states are non-overlapping in configuration space is a special property of the position representation. The same states might not be non-overlapping in an other representation. This is very important for the choice of beable since it is unclear how a pilot-wave model in which macroscopically distinct states are overlapping in the corresponding configuration space reproduces the quantum predictions \ldots Similar issues will arise when introducing field beables for quantum field theory, i.e. merely introducing some field beable which is distributed according to a quantum equilibrium distribution will not guarantee that the pilot-wave model reproduces the standard quantum predictions. \ldots

In the pilot-wave theory of de Broglie and Bohm for non-relativis\-tic quantum theory, two states representing a single particle localized at different regions in physical space are non-overlapping. With the field ontoloit isgy this is not true anymore: wavefunctionals describing a particle localized at different regions have significant overlap. Even stronger, we can show that any two wavefunctionals describing a single particle have significant overlap. \ldots

We think of a macroscopic system, like a measurement needle, not only as an approximately localized system, but also as a system composed of a large, but approximately fixed number of particles. Therefore it would be more interesting to consider wavefunctionals for such systems and see in which cases they might be non-overlapping. Maybe the above argument that shows that one-particle wavefunctionals always have significant overlap can be extended to wavefunctionals describing a large but fixed number of particles. \cite{Struyve}
\end{quote}

To solve this problem, we define a precise notion for the overlap of two functions, with the physical meaning of the probability that a configuration guided by one wave functional appears to be located in a region where the other wave functional dominates. Based on this definition, we find, in agreement with Struyve, that the overlap between orthogonal n-particle states in field theory is a constant, which depends only on the number of particles, but not on their locations in physical space. And, again in agreement with Struyve, we find that for one-particle states this overlap has the value $0.18169 (\pm 1)$, which is clearly not negligable.

To obtain the overlap between orthogonal $n$-particle states, we have to compute an integral in $2n$-dimensional space. We use Monte Carlo computation to obtain approximations of the overlap for small numbers of particles. A simple laptop has been sufficient to obtain reliable results for $n<20$. We find that the overlap decreases with the number of particles in a very fast, approximately exponential, way. For $n=16$ the overlap has been reduced to a value below $0.01\%$, and the (already unreliable) value for $n=20$ is below $0.001\%$. Thus, already for small numbers of particles we obtain overlaps which are negligible for most parts of human experience.

We also present some plausibility arguments that the almost exponential decrease of the overlap is not an accident for small particle numbers, but will continue for larger particle numbers as well.

Taken together, and taking into account that an exponential decrease of the overlap in the number of particles is much more than we need to establish the recovery of quantum predictions for macroscopic particle numbers, we conclude that the problem raised by Struyve does not prevent field beables from distinguishing macroscopically different states. Thus, pilot wave theories with field ontology are able to recover quantum field theory predictions without any problems.

\section{The measurement process in pilot wave theory}

Let's consider the general scheme of a quantum measurement in pilot wave theory. We restrict ourself in this paper to the simplest case of a measurement with two discrete eigenstates. This seems sufficient: As human beings, we can distinguish only finite numbers of different states, and, given that the number of particles in macroscopic states is much larger than the number of states we can distinguish, the number of the discrete states does not really matter. So assume we have a quantum system $S$ in the initial state $\psi(t_0,q_S)=a_0\psi_0^S(q_S) + a_1\psi_1^S(q_S)$, with $\langle\psi_0^S|\psi_1^S\rangle=0$. The measurement is an interaction with some measurement device, which measures the projection operator $|\psi_1^S\rangle\langle\psi_1^S|$ --- an operator with two eigenvalues, $1$ for $\psi_1$ and $0$ for all other states. For a simple toy model, the interaction Hamiltonian may be something like $|\psi_1^S\rangle\langle\psi_1^S|p_m$, where $p_m$ is the momentum operator of some quantum pointer variable, which is, initially, in some localized state $\psi^m_{0}(t_0,q_m)=e^{-q_m^2}$. After the interaction, we obtain some superpositional state
\begin{equation}
\psi(t_1,q_S,q_m) = a_0\psi_0^S(q_S)\psi^m_{0}(q_m) + a_1\psi_1^S(q_S)\psi^m_{1}(q_m)
\end{equation}
with $\psi^m_{1}(q_m)=e^{-(q_m-\Delta T)^2}$. Now pilot wave theory allows to define for this superpositional state the so-called conditional wave function
\begin{equation}
\psi_{cond}(t_1,q_S) = \psi(t_1,q_S,q_m),
\end{equation}
where $q_m$ is the actual state of the configuration of the measurement device. This conditional wave function gives at every moment of time the correct guiding equation. But in general, it does not follow an effective Schr\"{o}dinger equation. Only if $T$ is sufficiently large, so that the two wave functions $\psi^m_{0}(t_0,q_m)=e^{-q_m^2}$ and $\psi^m_{1}(q_m)=e^{-(q_m-\Delta T)^2}$ no longer overlap, and the measurement interaction has finished, the situation is different.If $q_m$ is located in the support of $\psi^m_{0}(T,q_m)$ (or $\psi^m_{1}(T,q_m)$), its location inside this support already does not matter: the conditional wave function will be $\psi_0^S(q_S)$ (resp. $\psi_1^S(q_S)$). Moreover, this conditional wave function will be also an effective wave function: The particle does no longer interact with the measurement device, thus, the effective future evolution of the wave function depends only on the system $S$. Moreover, the Hamilton operator is local in the configuration space, thus, the part of the wave function which has been omitted, because it is localized in a different region, does not influence the evolution. Thus, the reduced Hamilton operator acting on $\psi_0^S(q_S)$ (resp. $\psi_1^S(q_S)$) defines effectively the future evolution. The Born rule follows from quantum equilibrium: The value $q_m$ is located in the support of $\psi^m_{0}(T,q_m)$ (or $\psi^m_{1}(T,q_m)$), with probability $\abs{a_0}^2$ (resp. $\abs{a_1}^2$).

Unfortunately, our toy measurement theory is not enough: For one-dimensional wave packets $\psi^m_{i}(T,q_m)$, the probability that they meet again in some future is simply too large to be ignored. We need some macroscopic amplification of the measurement results to be sure that this cannot happen. But formally nothing changes: We obtain states of some classical measurement device $\psi^M_{i}(q_M)$, which play the same role as the $\psi^m_{i}(q_m)$. The difference is only that that these states are states in a much larger Hilbert space, described by a much larger set of variables $q^k_M$. Now, for these macroscopic states, it is already reasonable to expect that they don't overlap in all of the variables $q^k_M$, and that this property remains stable in time.

But there may be some other issues which prevent the functions $\psi^M_{i}(q_M)$ from having a non-significant overlap. In particular, Struyves argument can be reformulated as a particular counterexample for a pilot wave theory where the $\psi^M_{i}(q_M)$ always have significant overlap: This counterexample is one-particle theory formulated, in an artifical way, as a pilot wave theory with field beables. It appears that the overlap between one-particle states of field theory is always significant. Our computations below give a value of $0.18169 (\pm 1)$, which is far away from being insignificant.

Fortunately, this counterexample is not the physically relevant one. Beyond the fact that it does not describe macroscopic states, we need a special conservation law (of the number of particles) together with a very special initial value (only one particle) to obtain this failure. Nonetheless, this counterexample proves that it is important to look in more detail at the overlaps between macroscopic states, to be sure that everything is ok.

\section{The definition of the overlap}\label{definitions}

If two wave functions $\psi_0$, $\psi_1$ overlap or not is a well-defined notion: if the product $\psi_0(q)\psi_1(q)$ is zero everywhere, they don't overlap, else they overlap. But this precise definition is not helpful, because that never happens exactly. Even the typical example of a localized function --- the Gauss distribution $e^{-x^2}$ --- is non-zero everywhere. Therefore, one needs a more practicable definition of the overlap which allows for small, negligable overlaps, one which assigns some precise value $\rho(\psi_0|\psi_1)\le 1$ to the overlap. Then the choice of the $\varepsilon$ so that  $\rho(\psi_0|\psi_1)\le \varepsilon$ for an overlap to be negligible can be left to the particular application.

How this can be done? A natural way to make the notion of negligible overlap precise is to consider the error we have to make if we replace two given functions by exactly non-overlapping approximations. Thus, let's approximate the two wave functions $\psi_i(q)$, $0\le i\le 1$, by non-overlapping approximations $\psi_i \approx \widetilde{\psi}_i$. There is a quite natural way to do this, defined by the following rule:
\begin{equation}
 \widetilde{\psi}_i(q) =\begin{cases}
    0		& \text{if} \quad|\psi_i(q)|\le|\psi_{1-i}(q)|,\\
    \psi_i(q)	& \text{otherwise}.
  \end{cases}
\end{equation}
In other words, the only wave function which remains nonzero in a given point is that which has the maximal value. This is reasonable, because the error of approximating it by zero would be larger than for the other function. Given this definition, it is obvious that the approximations $\widetilde{\psi}_i$ do not overlap --- in each point, by construction, at most one of them is nonzero.

What is the error made by this approximation? It is defined in a natural way by the \L-norm of the differences $\ltwo{\psi_i - \widetilde{\psi}_i}$ between the original wavefunctionals and their approximations. For the overlap of $\psi_0$ relative to $\psi_1$, this is
\begin{equation}\label{overlapdef}
 \rho(\psi_0|\psi_1) = \int\limits_{\abs{\psi_0(q)}\le\abs{\psi_1(q)}}\abs{\psi_0(q)}^2 d^Nq.
\end{equation}
For the corresponding overlap $\rho(\psi_1|\psi_0)$ of $\psi_1$ relative to $\psi_0$ the formula is similar.

Above parts of the overlap have a simple probability interpretation: $\rho(\psi_i|\psi_j)$ defines the probability of a particle guided by the wavefunction $\psi_i$ to be localized, in quantum equilibrium, in a region where $\abs{\psi_j}$ is greater than $\abs{\psi_i}$. If we exclude the degenerated case $|\psi_i(q)|=|\psi_{1-i}(q)|$, we have the inequality $\rho(\psi_0|\psi_1)+\rho(\psi_1|\psi_0)\le 1$ (in the case of equality the sum may go up to $2$), thus, a small overlap will be characterized by $\rho(\psi_0|\psi_1)+\rho(\psi_1|\psi_0)\ll 1$. The actual meaning of ``$\ll 1$'' depends of course on the accuracy required in the particular application. The probability interpretation allows to apply it in the equivalence proof: If the overlap is small enough, this can be translated into a small enough probability that something fails with the equivalence between pilot wave theory and the predictions of quantum theory (taken together with the assumption that the measurement results have been already sufficiently amplified).

\section{The overlap between n-particle states in field theory}\label{nnoverlap}

In pilot wave theories with field ontology the configuration space $\Phi$ is a space of functions. We can restrict ourself to the simplest case of a real scalar field $\Phi=\{\phi(x):\R^3\to\R\}$. For the purpose of this paper, the functional-analytical problems which appear in field theories play no role at all. We can imagine the continuous space variable $x\in\R^3$ to be replaced by some lattice approximation, so that $x$ takes only discrete values on some lattice $L\subset\Z^3$ with $N$ points. Then, $\phi(x)$ is simply another notation, which replaces $q^i$, with $x$ playing the role of the index $i$. The wave functions in field theory are, now, functions on the space of functions, which are called wavefunctionals, and a large $\Psi$ is used to denote them.

The field theory states we consider in this paper are the following: First, the vacuum state, which is described by a Gaussian distribution
\begin{equation}\label{def:vacuum}
 \Psi_{vac}(\phi) =  \frac{1}{\sqrt{\pi}^N}\quad  e^{-\frac{1}{2}\sum_{x=1}^N \phi(x)^2}
\end{equation}
Then, we have one-particle states of the field theory. They are defined by some normalized field configuration $\psi(.)\in\Phi$. The wave functional of the one-particle state of the field $\psi$ is defined in the following way:
\begin{equation}
 \Psi_{\psi}(\phi) = \sqrt{2}\langle\psi,\phi\rangle \cdot \Psi_{vac}(\phi)
\end{equation}
We need also n-particle states. But we consider here only a special type of n-particle states -- states which are defined by $n$ different orthogonal states $\psi_k$, while for a general n-particle state the states $\psi_k$ don't need to be orthogonal. This type seems to be the most interesting one in our case: The argument has been about states with an approximately fixed particle number, which are usually fermions, and for fermionic n-particle states the $\psi_k$ have to be orthogonal. Otherwise, the $\psi_k$ may be arbitrary field configurations. The n-particle states can be obtained in the same way as the one-particle state:
\begin{equation}\label{nparticledef}
\Psi_n(\phi)=\Psi_{\psi_1,\ldots\psi_n}(\phi)=\prod\limits_{k=1}^n\sqrt{2}\langle\psi_k,\phi\rangle \cdot \Psi_{vac}(\phi).
\end{equation}

The problem posed by Struyve \cite{Struyve} is that the overlap between one-particle states in field theory is always significant, even if the field configurations themself do not overlap at all in physical space. This suggests that a similar problem may appear for larger particle numbers as well. The aim of this section is to find a formula which allows to compute this overlap, as defined by equation \eqref{overlapdef}, for n-particle states.

A key observation for the evaluation of this integral is that we are free to choose a basis in the space of field variables $\phi(x)\in \Phi\cong \mathcal{L}^2(\R^3,\C)$. Thus, instead of the $\delta$-function basis where the coordinates of function $\phi\in\Phi$ are the function values $\phi(x), x\in L$, we can use some other set of basic fields $\varepsilon_i(x)\in\Phi$ so that an arbitrary field is defined by the set $\{\phi_i\}$ of components in the decomposition
\begin{equation}
 \phi(x) = \sum_i \phi_i \varepsilon_i(x).
\end{equation} 

This should not be mingled with the choice of a ``preferred basis'' in the Hilbert space of wave functionals $\H$, which amounts to the choice of the configuration space $\Phi$. Once the space $\Phi$ of field configurations has been chosen as the configuration space or space of beables, the Hilbert space basis is already fixed, and the Hilbert space of wave functionals is represented by square-integrable functionals on $\Phi$ as $\H\cong\L(\Phi,\C)$.

Thus, we are free to change the basis in $\Phi$. We use this freedom to compute the overlap integral between two orthogonal n-particle states $\Psi_n^0(\phi)$, $ \Psi_n^1(\phi)$ defined by
\begin{align}
  \Psi_n^0(\phi) &= \prod\limits_{k=1}^n\sqrt{2}\langle\psi^0_k,\phi\rangle \cdot \Psi_{vac}(\phi),\\
  \Psi_n^1(\phi) &= \prod\limits_{k=1}^n\sqrt{2}\langle\psi^1_k,\phi\rangle \cdot \Psi_{vac}(\phi),
\end{align}
using a basic containing all the $2n$ fields $\psi^0_k(x)$, $\psi^1_k(x)$ as the first $2n$ basis vectors, so that $\phi_{2k-1}=\langle\psi^0_k,\phi\rangle$, $\phi_{2k}=\langle\psi^1_k,\phi\rangle$. Once we consider here only the case where all these $2n$ fields are orthonormal, this is a possible choice. To compute the functional integral
\begin{equation}
\rho_n=\rho( \Psi_n^0, \Psi_n^1) = \int\limits_{\abs{\Psi_n^0(\phi)}\le\abs{\Psi_n^1(\phi)}} \abs{\Psi_n^0(\phi)}^2 \prod_{i=1}^{N} d\phi_i
\end{equation}
we observe at first that, whatever the orthonormal basis in $\Phi$, the vacuum state will not change it's form -- it remains a product of normalized Gauss distributions of each of the variables:
\begin{equation}\label{def:vacuum1}
 \Psi_{vac}(\phi) = \prod\limits_{x=1}^N (\frac{1}{\sqrt{\pi}} e^{-\frac{1}{2}\phi(x)^2})
		  = \prod\limits_{i=1}^N (\frac{1}{\sqrt{\pi}} e^{-\frac{1}{2}\phi_i^2})
\end{equation}
Then, all other constituents of the integral -- the domain of integration defined by $\abs{\Psi_n^0(\phi)}<\abs{\Psi_n^1(\phi)}$ and the factor distinguishing the $\Psi_n^{0/1}$ from $\Psi_{vac}$ -- depend only on the first $2n$ variables. Thus, the integral over the remaining $N-2n$ degrees of freedom is a simple appropriately normalized Gauss integral
\begin{equation}
\prod\limits_{i=2n+1}^N \int \frac{1}{\pi} e^{-\phi_i^2} d\phi_i = 1
\end{equation}
Thus, the integral over these $N-2n$ degrees of freedom can be taken explicitly and gives the trivial factor $1$.

What remains from our $N$-dimensional (in the limit infinite-dimensional) integral is therefore an integral over the first $2n$ variables:
\begin{equation}\label{rhonn}
\rho_n=\frac{2^n}{\pi^n} \int\limits_{\prod\limits_k\phi_{2k-1}^2\le\prod\limits_k\phi_{2k}^2}
	(\prod\limits_k\phi_{2k-1}^2 )e^{-\sum\limits_k(\phi_{2k-1}^2+\phi_{2k}^2)} \prod\limits_k \d\phi_{2k-1} \d\phi_{2k}.
\end{equation}

The field-theoretic limit $N\to\infty$ is now a triviality. A consequence is that the overlap between two orthogonal depends only on the number $n$ of the particles involved. The same holds in other field theories, where the coordinate space $X\cong \R^3$ may be replaced by some $\R^3\times \Sigma$ where $\Sigma$ describes other degrees of freedom of the field like the spin: Indeed, nothing in our consideration depends on it. All we need is orthogonality of the $2n$ states in the space of field configurations $\Phi$. Given the orthogonality, the overlap reduces to the integral \eqref{rhonn}, which depends only on $n$.

In particular, the overlap of the field configurations in the coordinate space $X$ does not matter at all for the overlap of the corresponding wave functionals in field space. Locality becomes important in field theories only if we consider the dynamics, where, in case of a lattice as well as in the case of field theory, we can restrict the interactions to nodes which are neighbours in the usual coordinate space $X$.

\section{Monte Carlo computation results for small particle numbers}\label{nncomputation}

\begin{figure}
\centerline{\includegraphics[angle=0,width=\textwidth]{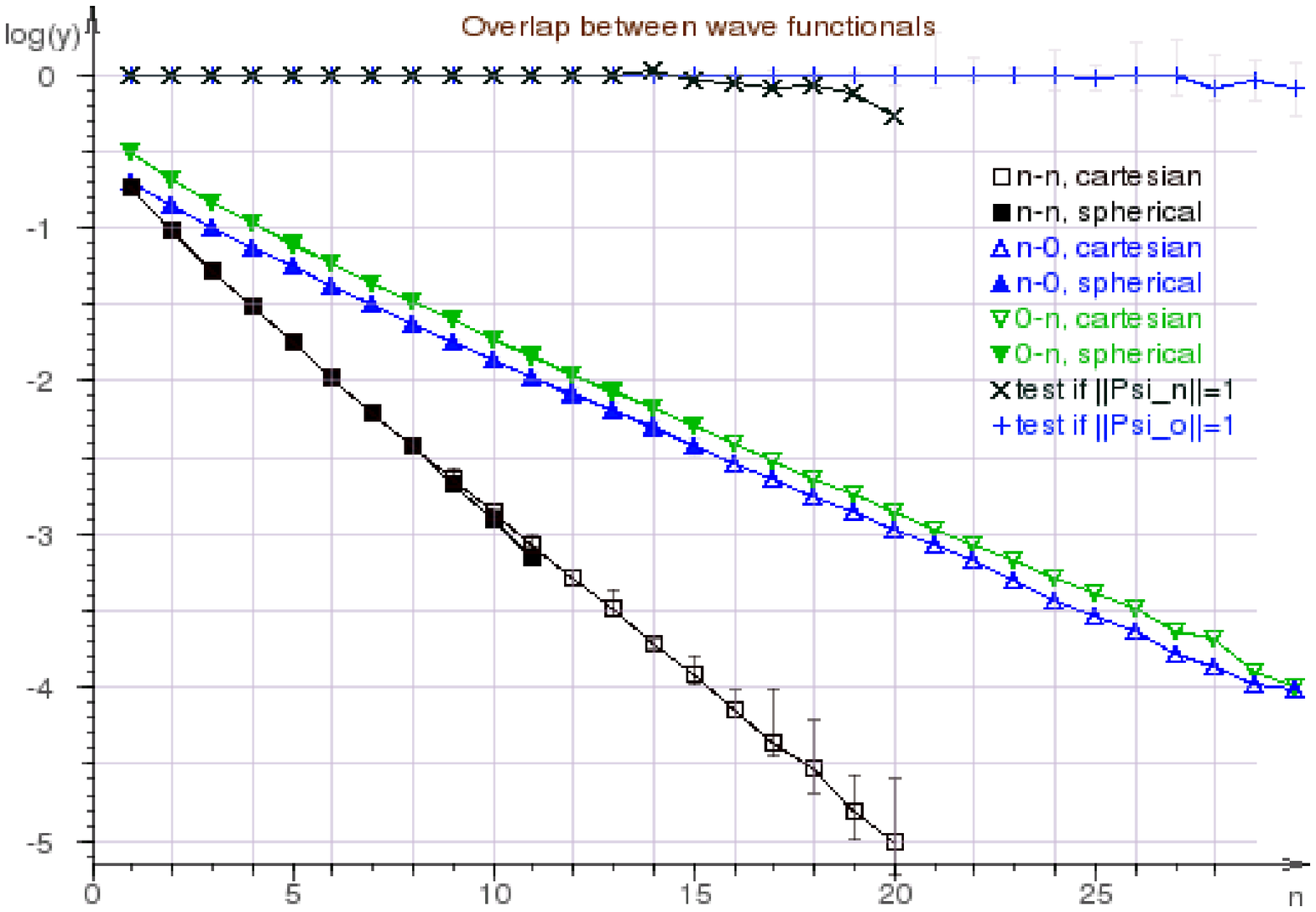}}
\caption{\label{fig1} Logarithmic plot for the Monte Carlo simulation results}
The curves marked as $n-n$ (boxes) show the overlap $\rho_n$ defined by \eqref{rhonn} between two orthogonal states with n particles in each state. The curves marked as $n-0$ and $0-n$ (triangles) show the overlaps $\rho(\Psi_n|\Psi_{vac})$ resp. $\rho(\Psi_{vac}|\Psi_n)$ between an n-particle state $\Psi_n$ and the vacuum state $\Psi_{vac}$. The upper curves show $\ltwo{\Psi_n}$ resp. $\ltwo{\Psi_{vac}}$ for the point sets used in the computation (which should be $1$, that means, $0$ in the logarithmic plot). The results for point distributions based on spherical coordinates (full boxes resp. triangles) are less reliable and more expensive, and therefore have been computed only for smaller particle numbers.
\end{figure}

These constants seem sufficiently interesting to compute some first of them numerically. An appropriate method for this purpose seems to be a Monte Carlo approximation: One does not need very complex software, and it appears that a simple laptop is sufficient today to run rather large Monte Carlo computations, sufficient to obtain more or less reliable results for at least the first 16 particles, and sufficient to obtain some hints about their dependence on the number of particles.

The results can be seen in figure \ref{fig1}. Because of the fast decrease, it was necessary to use a logarithmic plot to show all the values which can be reliably computed. The first value is $0.18169(\pm 1)$. Already for $n=2$ this is reduced to $0.0947(\pm 1)$. The $1\%$ level is reached for $n=6$ with $0.0104(\pm 3)$. For $n=11$, $9\cdot 10^{-4}(\pm 2)$, we are already below $0.1\%$, and for $n=16$ below $0.01\%$. The last value we have computed, and which seems already unreliable given our reliability criteria, is $n=20$, and is below $10^{-5}$.

A problem of Monte Carlo is that it always gives some result, without indication that it may be unreliable. The main reason for Monte Carlo to become unreliable are sharply localized functions. The wave functions we have to consider become, with higher dimension, more and more localized. In such unreliable situations, most computations will underestimate the value (with not enough points in the peaks), while seldom the result will be far too large. Thus, there is a danger of underestimating the overlap because of the numerical problems of the Monte Carlo method. So we need some estimates of the accuracy of the computation.

One simple way is to subdivide the whole set of points into packages and to compare the results for the packages. We have used 20 packages, and taken the maximum and minimum over these subpackages as error bars.

As another reliability test, we have used the fact that the integral \eqref{rhonn} without the characteristic function should give the norm, thus, exactly $1$. Thus, we have computed, for the same set of nodes, this integral too, to see if it recovers the $1$ with sufficient accuracy (two upper curves in fig. \ref{fig1}).

Then we have used two different ways to obtain random point sets --- the first one based on cartesian coordinates, the second one using spherical coordinates. While it has appeared to be surprisingly simple to implement a random number generator based on $n$-dimensional spherical coordinates, the results for spherical coordinates have been clearly worse. This can be easily explained: In spherical coordinates, the nodes are more dense in the ``polar'' regions. While this is, of course, incorporated in the measure, it nonetheless influences the accuracy of the computation. In our particular overlap integral, the polar regions are regions where nothing interesting happens. Instead, for the random point sets based on cartesian coordinates, the point set is more dense in the diagonal regions. And, in case of our overlap integral, this is the region which is important, in particular because the maxima of the functions are located in the diagonal directions. Moreover the computation needs more time than in the case of cartesian coordinates. Therefore we have been unable to compute them for larger numbers of particles. Nonetheless, for the region where we have computed them, and where they give accurate results using the previously mentioned criteria, they recover the results of the computations for cartesian coordinates.

We have computed also, for comparison, the overlaps between the $n$-particle states $\Psi_n$ and the vacuum state $\Psi_{vac}$. In this case, the two overlap probabilities $\rho(\Psi_n|\Psi_{vac})$ and $\rho(\Psi_{vac}|\Psi_{n})$ are different. These integrals have the advantage that one needs only $n$-dimensional space (instead of $2n$ for \eqref{rhonn}) to compute them. This gives reliable results for higher particle numbers (as can be seen from the test if $\ltwo{\Psi_n}$ resp. $\ltwo{\Psi_{vac}}$ give $1$ in the two upper curves of fig. \ref{fig1}).

The logarithmic plot shows approximately straight lines, indicating an approximately exponential decrease. The average value for the overlap between n-particle states $\rho_n$ decreases approximately like $1.7^{-n}$. Initially the rate drops slightly, from $1.9^{-n}$, between the first two values, down to around $1.62^{-n}$ for larger values, but this trend becomes less important for larger n, where the logarithmic plot looks much more straight. The two curves for $\rho(\Psi_n|\Psi_{vac})$ and $\rho(\Psi_{vac}|\Psi_{n})$ show a similar pattern, in spherical as well as cartesian coordinates. Therefore we think that this effect --- the initial drop of the rate of decrease, as well as that this effect becomes less significant later --- is not a numerical artefact, but a real property of this type of overlap integrals.

We cannot decide with our methods if this decrease is a low-dimensional effect without relevance for larger $n$, or an indication that the decrease is not exponential but subexponential (say, something like $c^{-\sqrt{n}}$). Fortunately, this question is not really important. Given the large numbers of particles in macroscopic states, even a much smaller decrease, say, a polynomial one with a small degree like $\sqrt{n}^{-1}$, would be sufficient to obtain a negligible overlap.

\section{Additional plausibility arguments}

Of course, computations for small numbers do not prove anything in a strong mathematical sense. But what we compute is a quite simple integral \eqref{rhonn}, sufficiently simple that it seems reasonable to assume that one can extrapolate it's behaviour for small $n$ to larger $n$.

Moreover, we do not need very much extrapolation: Last but not least, the significant one-particle overlap of 18\% has been already reduced to 1\% for six particles, 0.1\% for eleven particles, and 0.01\% for sixteen particles. Such a small overlap is already negligible for most of human experience.

But one can add yet another hint that this decrease will continue: The distance between the maxima of the wave functions can be easily computed, and, as well, depends only on the number of particles. Indeed, the maxima of $\phi^2e^{-\phi^2}$ are located at $\pm 1$. Therefore n-particle states in \eqref{rhonn}, which are product states $\prod \phi_{2k/2k-1}^2e^{-\phi_{2k-1}^2+\phi_{2k}^2}$, have their maxima at $(\pm 1,0,\ldots,\pm 1,0)$ resp. $(0,\pm 1,\ldots,0,\pm 1)$. This gives the distance $\sqrt{2n}$ between the maxima of n-particle states in \eqref{rhonn}
maximum of the n-particle state and the vacuum state, and the distance $\sqrt{n}$ between the maxima of $\Psi_n$ and $\Psi_{vac}$.

If we additionally assume that the n-particle states are localized in a way comparable to that of the vacuum state, we have to expect that the overlap decreases with distance $\Delta$ as $e^{-\Delta^2}$. This suggests, for $\Delta = \sqrt{2n}$, an exponential decrease, in nice correspondence to our numerical results.

Note that this argument works in a particular nice way for the overlap $\rho(\Psi_{vac}|\Psi_{n})$ which we have computed too. In this case, the first function --- the vacuum state --- decreases exactly like $e^{-\Delta^2}$. Given that the qualitative behaviour of this function is very similar to that of $\rho_n$ (namely approximately exponential decrease with a small decrease in the exponent, which seems to be relevant only for small numbers $n$), it seems plausible that the qualitative behaviour of these two functions remains similar.

\section{Conclusions}

We have considered Struyve's problem if the overlap between orthogonal $n$-particle states is significant. For this purpose, we have proposed a precise definition of the overlap with natural probability interpretation. Based on this definition, we have computed the overlap between orthogonal states consisting of $n$ orthogonal particles in field theory for $n\le 20$ using Monte Carlo approximation. We have found an approximately exponential decrease of the overlap, starting from the non-negligible overlap of $0.18169$ between two one-particle states down to below $10^{-4}$ for $n=16$ particles in each state.

We have also found an additional plausibility argument that a similar approximately exponential decrease will continue for larger $n$: The distance between the maxima of $n$-particle states increases like $\sqrt{2n}$.

What is necessary to recover quantum predictions is much less than the approximately exponential decrease we have found plausible: All we need is that the overlap becomes negligible for macroscopic particle numbers. Given that a number as small as $16$ already reduces the overlap from $18\%$ to below $0.01\%$, which is already negligible for most of our human experience, our observations seem sufficient: There is no reason to doubt that the overlap between macroscopic states will be negligible.

As a consequence, the equivalence proof between standard quantum theory and pilot wave theory in quantum equilibrium works even for the case of field beables. The known pilot wave theories with field ontologies (see \cite{Struyve} for an overview) are therefore sufficient to recover the empirical predictions of quantum field theories.

This certainly does not mean that the situation for pilot wave field theories is satisfactory: The existing proposals for field theories may be criticized for a lot of different metaphysical reasons. In particular, there is no nice field ontology for fermions, and to consider only bosons, while sufficient to recover the empirical predictions, is certainly ugly. In this context, the condensed matter interpretation for the SM proposed by the author in \cite{clm} may be of interest: The quantization scheme for fermions used there is a variant of canonical quantization, thus, allows to apply the standard scheme of pilot wave theory to obtain a pilot wave version of this model.

But, whatever the problems of pilot wave field theories in the domain of quantum field theories, the ability to recover the empirical prediction of quantum field theory is in general\footnote{Of course we have shown this only for the particular case of a simple real field. For other particular theories the situation may be different. I would like to thank one of the referees for pointing me to section 9.1.3 added in the second version of \cite{Struyve}, which contains an interesting new argument against empirical viability of a particular theory with field ontology. At a first look it seems based on the assumption that the beables have to observables, which is not necessary in the variant of the equivalence proof discussed in \cite{overlap2}. But this clearly deserves future consideration.} not among them. This is an important fact, not only for some particular version of pilot wave interpretations. It proves the viability of classical physical principles like classical realism, determinism, and classical causality in the domain of relativistic field theory: Pilot wave theories have all these properties, and the existence of empirically viable pilot wave field theories proves that these general classical principles are viable even in the domain of relativistic quantum theory.

\end{document}